\documentclass[11pt,a4paper]{article}
\usepackage{jheppub}
\usepackage[T1]{fontenc}
\usepackage{amsmath}
\usepackage{amssymb}
\usepackage{mathrsfs}
\usepackage{graphicx}
\usepackage{float}
\usepackage{bm}

\title{\boldmath Transverse momentum spectrum of dilepton pair in the unpolarized $\pi^-N$ Drell-Yan process within TMD factorization}
\author[]{Xiaoyu Wang$^{a}$, Zhun Lu$^{a}$,}
\affiliation[]{$^{a}$School of Physics, Southeast University, Nanjing, Jiangsu 211189, China}
\author[]{Ivan Schmidt$^{b}$}
\affiliation[]{$^{b}$Departamento de F\'\i sica, Universidad T\'ecnica Federico Santa Mar\'\i a, and
Centro Cient\'ifico-Tecnol\'ogico de Valpara\'iso,
Casilla 110-V, Valpara\'\i so, Chile}

\abstract{We study the transverse momentum spectrum of dilepton produced in the unpolarized $\pi^- N$ Drell-Yan process, using transverse momentum dependent factorization up to next-to-logarithmic order of QCD. We extract the nonperturbative Sudakov form factor for the pion in the evolution formalism of the unpolarized TMD distribution function, by fitting the experimental data collected by the E615 Collaboration at Fermilab. With the extracted Sudakov factor, we calculate the normalized differential cross section with respect to transverse momentum of the dimuon and compare it with the recent measurement by the COMPASS Collaboration.}

\begin{document}
\maketitle
\flushbottom
\section{Introduction}
After the first observation of the $\mu^+ \mu^-$ lepton pairs produced in $p\,N$ collisions~\cite{Christenson:1970um}, the process was interpreted as a quark and an antiquark from each initial hadron annihilating into a virtual photon, which in turn decays into a lepton pair~\cite{Drell:1970wh}. This explanation makes the process an ideal tool in order to explore the internal structure of both the beam and target hadrons. Since then, a wide range of studies on this (Drell-Yan) process have been carried out.
In particular, the $\pi N$ Drell-Yan process has the unique capability to pin down the partonic structure of the pion,
which is an unstable particle, it therefore cannot serve as a target in deep inelastic scattering processes.
Several pion induced experiments have been carried out, such as NA10 at CERN~\cite{Bordalo:1987cs,Betev:1985pf,Falciano:1986wk,Guanziroli:1987rp}, E615~\cite{Conway:1989fs}, E444~\cite{Palestini:1985zc} and E537~\cite{Anassontzis:1987hk} at Fermilab. These measurements have provided plenty of data, which have been used to constrain considerably the pion distribution function.
A new pion-induced Drell-Yan program was also proposed to be executed at the COMPASS facility~\cite{Gautheron:2010wva}.
In particular, the COMPASS collaboration measured for the first time the transverse-spin-dependent azimuthal asymmetries in the $\pi^- N$ Drell-Yan process, using a high-intensity $\pi$ beam of 190 GeV colliding on a transversely polarized NH$_3$ proton target~\cite{Aghasyan:2017jop}.

One of the main goals of the COMPASS Drell-Yan program is to measure the Sivers asymmetry or the Sivers function with high precision. The Sivers function is a transverse momentum dependent (TMD) parton distribution function (PDF) describing the asymmetric density of unpolarized quarks inside a transversely polarized nucleon. Remarkably, QCD predicts that the sign of the Sivers function changes in SIDIS with respect to the Drell-Yan process~\cite{Brodsky:2002cx,Brodsky:2002rv,Collins:2002kn}. The verification of this sign change~\cite{Anselmino:2009st,Kang:2009bp,Peng:2014hta,Echevarria:2014xaa,Huang:2015vpy,Anselmino:2016uie} is one of the most fundamental tests of our understanding of the QCD dynamics and the factorization scheme, and it is also the main pursue of the existing and future Drell-Yan facilities~\cite{Aghasyan:2017jop,Gautheron:2010wva,Fermilab1,Fermilab2,ANDY,Adamczyk:2015gyk}. The advantage of the $\pi\,N$ Drell-Yan measurement at COMPASS is that almost the same setup~\cite{Aghasyan:2017jop,Adolph:2016dvl} is used in SIDIS and Drell-Yan, which may reduced the uncertainty in the extraction of the Sivers function.

In order to quantitatively understand the Sivers asymmetry in the $\pi N$ Drell-Yan process at COMPASS, a first step is to know with high accuracy the differential cross-section of the same process for an unpolarized nucleon, since the spin-averaged cross-section is what appears in the denominator of the asymmetry definition. This motivates us to study the unpolarized $\pi N$ Drell-Yan process in a comprehensive way.
Since most of the data accumulated by the COMPASS collaboration are at low transverse momentum (of the dilepton $q_\perp$), a natural choice for the analysis framework is TMD factorization, which is valid in the region where $q_\perp$ is much smaller than the hard scale $Q$.
The TMD factorization has been widely applied to various high energy processes, such as the SIDIS~\cite{Collins:1981uk,Collins:2011zzd,Ji:2004wu,Aybat:2011zv,Collins:2012uy,Ji:2002aa,Echevarria:2012pw},
$e^+ e^-$ annihilation~\cite{Collins:2011zzd,Pitonyak:2013dsu,Boer:2008fr}, Drell-Yan~\cite{Collins:2011zzd,Boer:2006eq,Arnold:2008kf} and W/Z production in hadron collision~\cite{Collins:2011zzd,Collins:1984kg,Lambertsen:2016wgj}.
From the point of view of TMD factorization~\cite{Collins:1981uk,Collins:1984kg,Collins:2011zzd,Ji:2004xq}, physical observables can be written as convolutions of a factor related to hard scattering and well-defined TMD distribution functions or fragmentation functions. We will calculate the cross-section of the process up to the next-to-leading logarithm (NLL) and also take into account the evolution effects of the TMD distributions.

In contrast to what is done in the collinear PDFs case, the evolution of TMDs is usually performed in $b$ space~\cite{Collins:1984kg,Collins:2011zzd}, which is conjugate to the transverse momentum $\bm{k}_\perp$. Moreover, the TMD PDFs depend on two energy scales. One is related to the corresponding collinear PDFs and the other is related to the definition of the TMD PDFs. After solving the evolution equations, the TMDs at fixed energy scale can be expressed as a convolution of their collinear counterparts and perturbatively calculable coefficients,
and the evolution from one energy scale to another energy scale is included in the exponential factor of the so-called Sudakov-like form factors~\cite{Collins:1984kg,Collins:2011zzd,Collins:1999dz}. The Sudakov form factor can be separated into a perturbatively calculable part and a nonperturbative part, the later one can be fitted to experimental data.
In this work, we perform the first extraction of the nonperturbative Sudakov form factor for the unpolarized TMD PDF $f_1$ of pion meson in the $\pi^-P$ Drell-Yan process, which sheds light on the partonic structure of the pi meson.


The rest of the paper is organized as follows.
In Sec.~\ref{Sec.Framework} we set up the necessary theoretical framework of the TMD factorization and evolution effects in the unpolarized $\pi^-N$ Drell-Yan process. We also propose a prescription for the nonperturbative Sudakov form factor in the case of the spin-independent pion distributions. In Sec.~\ref{Sec.Numerical} we calculate numerically the transverse momentum dependent differential cross section, based on the framework given in Sec.~\ref{Sec.Framework}, and we fit the nonperturbative Sudakov form
factor using the $q_\perp$-dependent experimental data obtained by the E615 collaboration at Fermilab~\cite{Conway:1989fs}. In Sec.~\ref{Sec:COMPASS} we present the prediction for the $q_\perp$ spectrum of the dilepton pair produced in the $\pi^- N$ collision at the COMPASS kinematics, applying the extracted Sudakov form factor obtained in Sec.~\ref{Sec.Numerical}. In Sec.~\ref{Sec.conclusion} we summarize the results of the paper and give some conclusions.

\section{Framework of unpolarized Drell-Yan process in TMD factorization}

\label{Sec.Framework}
In this section we will set up the necessary framework of TMD factorization, following the procedure presented in Ref.~\cite{Collins:2011zzd}, in order to obtain the differential cross section for unpolarized Drell-Yan process considering TMD evolution effects.

\subsection{Kinematics and general formula for Drell-Yan Process}
The process under study is the unpolarized $\pi^- P$ Drell-Yan process
\begin{equation}
\label{eq:DYprocess}
\pi(P_\pi)+p(P_p)\longrightarrow \gamma^*(q)+X \longrightarrow l^+(\ell)+l^-(\ell')+X,
\end{equation}
where $P_\pi$, $P_p$ and $q$ denote the momenta of the $\pi^-$ meson, the proton and the virtual photon, respectively. In contrast to the SIDIS process, in the Drell-Yan process $q$ is a time-like vector, namely, $Q^2=q^2>0$, which is the invariant mass square of the lepton pair. In order to express the differential cross section of the process, we define the following kinematical variables
\begin{align}
&s=(P_{\pi}+P_p)^2,\quad x_\pi=\frac{Q^2}{2P_\pi\cdot q},\quad x_p=\frac{Q^2}{2P_p\cdot q},\nonumber\\
&x_F=2q_L/s=x_\pi-x_p,\quad\tau=Q^2/s=x_\pi x_p,\quad y=\frac{1}{2}\mathrm{ln}\frac{q^+}{q^-}=\frac{1}{2}\mathrm{ln}\frac{x_\pi}{x_p},
\end{align}
where $s$ is the center-of-mass energy squared; $x_\pi$ and $x_p$ are the light-front momentum fraction carried by the annihilating quarks in both the $\pi^-$ and the proton, respectively; $q_L$ is the longitudinal momentum of the virtual photon in the c.m. frame of the incident hadrons; $x_F$ is the Feynman $x$ variable, which corresponds to the longitudinal momentum fraction carried by the lepton pair; and $y$ is the rapidity of the lepton pair. Thus, $x_\pi$ and $x_p$ can be expressed as functions of $x_F$, $\tau$ and of $y$, $\tau$
\begin{align}
x_{\pi/p}=\frac{\pm x_F+\sqrt{x_F^2+4\tau}}{2},\quad x_{\pi/p}=\sqrt{\tau} e^{\pm y}.
\end{align}

Since it is much more convenient to solve the TMD evolution equations in coordinate space~($\bm{b}$-space) than in the transverse momentum space $\bm{q}_\perp$, where $\bm{b}$ is conjugate to $\bm{q}_\perp$ via Fourier Transformation, the differential cross section is usually expressed as a $b-$dependent function, formulated in TMD factorization. Later on it can be transformed back to transverse momentum space to represent the experimental observables. Thus, the differential cross section for the unpolarized $\pi^-$-proton Drell-Yan process described in (\ref{eq:DYprocess}) has the form~\cite{Collins:1984kg}
\begin{equation}
\label{eq:dsigma}
\frac{d^4\sigma}{dQ^2dyd^2\bm{q}_{\perp}}=\sigma_0\int \frac{d^2b}{(2\pi)^2}e^{i\vec{\bm{q}}_{\perp}\cdot \vec{\bm{b}}}\widetilde{W}_{UU}(Q;b)+Y_{UU}(Q,q_{\perp})
\end{equation}
where $\sigma_0=\frac{4\pi\alpha_{em}^2}{3N_CsQ^2}$ is the cross section at tree level. Hereafter, we use the tilde to denote $b-$space terms. The structure function $\widetilde{W}(Q;b)$ contains all-order resummation results and dominates at low $q_{\perp}\ll Q$ value, while the term $Y_{UU}$ provides the necessary correction at moderate $q_{\perp}\sim Q$ values. In this work we will neglect the $Y$-term, which means that we will only consider the first term on the r.h.s of Eq.~(\ref{eq:dsigma}).

In general, TMD factorization~\cite{Collins:2011zzd} aims at separating well defined TMD distributions, such that they can be used in different processes, and including the scheme/process dependence in the corresponding hard factors. Thus, $\widetilde{W}(Q;b)$ can be expressed as ~\cite{Prokudin:2015ysa}
\begin{align}
\label{eq:WUU_1}
\widetilde{W}_{UU}(Q;b)=H_{UU}(Q;\mu) \sum_{q,\bar{q}}e_q^2\tilde{f}_{1\, \bar{q}/\pi}^\mathrm{sub}(x_\pi,b;\mu,\zeta_F)
\tilde{f}_{1\,q/p}^\mathrm{sub}(x_p,b;\mu,\zeta_F),
\end{align}
where $\tilde{f}_{q/H}^\mathrm{sub}$ is the subtracted distribution function in $b-$space with the superscript "sub" representing the fact that the soft factor is subtracted, $H_{UU}(Q;\mu)$ is the factor associated with hard scattering, $\mu$ is renormalization scale in the case of collinear PDFs, and $\zeta_F$ denotes an energy scale related to the cutoff of the TMD distributions. We note that the distribution function $\tilde{f}_{1\,q/H}^\mathrm{sub}$ is universal, since it can be used in different processes. However, the way to subtract the soft factor in the distribution function depends on the scheme to regulate the light-cone singularity in the
TMD definition\cite{Collins:1981uk}, whereas the hard factor $H_{UU}(Q;\mu)$ is also scheme-dependent.
In the literature, two different schemes are usually used: the Collins-11 scheme~\cite{Collins:2011zzd} and the Ji-Ma-Yuan scheme~\cite{Ji:2004wu,Ji:2004xq}. If we perform a Fourier Transformation on $\tilde{f}_{1\,q/H}^\mathrm{sub}(x,b;\mu,\zeta_F)$, we obtain the distribution function in the transverse momentum space $f_{1\,q/H}(x,k_\perp;\mu,\zeta_F)$, which contains the information about the probability of finding a quark with collinear momentum fraction $x$ and transverse momentum $k_\perp$.

\subsection{TMD evolution and Sudakov form factor}
The TMD evolution for the $\zeta_F$ dependence of TMD PDFs is encoded in a Collins-Soper~(CS)~\cite{Collins:2011zzd} equation through
\begin{align}
\frac{\partial\ \mathrm{ln} \tilde{f}_1(x,b;\mu,\zeta_F)}{\partial\ \sqrt{\zeta_F}}=\tilde{K}(b;\mu),
\end{align}
while the $\mu$ dependence is derived from the renormalization group equation as
\begin{align}
&\frac{d\ \tilde{K}}{d\ \mathrm{ln}\mu}=-\gamma_K(\alpha_s(\mu)),\\
&\frac{d\ \mathrm{ln} \tilde{f}_1(x,b;\mu,\zeta_F)}
{d\ \mathrm{ln}\mu}=\gamma_F(\alpha_s(\mu);{\zeta^2_F\over\mu^2}),
\end{align}
where $\tilde{K}$ is the CS evolution kernel, and $\gamma_K$ and $\gamma_F$ the anomalous dimensions. The solutions of these evolution equations are studied in detail in Ref.~\cite{Collins:2011zzd}. Here, we will only discuss the final result.
The overall solution structure for $\tilde{f}_1(x,b;\mu,\zeta_F)$ is the same as that for the
Sudakov form factor. Namely, the energy evolution of TMDs from an initial energy $\mu$ to another energy $Q$ is encoded in the Sudakov-like form factor $S$ through the exponential form $\mathrm{exp}(-S)$\footnote{Hereafter, we set $\mu=\sqrt{\zeta_F}=Q$ and express $f(x,b;\mu=Q,\zeta_F=Q^2)$ as $f(x,b;Q)$ for simplicity.}
\begin{equation}
\tilde{f}(x,b,Q)=\mathcal{F}\times e^{-S}\times \tilde{f}(x,b,\mu).  \label{eq:f}
\end{equation}
Here, $\mathcal{F}$ is the hard factor, which depends on the scheme that we choose.

The $b$-dependence can determine the transverse momentum dependence of experimental observable through Fourier Transformation, which makes understanding the $b$-dependence become quite important. In the small $b$ region, the dependence is perturbatively calculable,  while at large $b$  the dependence turns nonperturbative and should be obtained from experiment data.
To combine the perturbative information at small $b$ with the nonperturbative part at large $b$, a matching procedure must be introduced with a parameter $b_{\mathrm{max}}$ serving as boundary between the perturbative and nonperturbative regions.
A $b$-dependent function $b_\ast$ is defined, which has the property $b_\ast\approx b$ at low values of $b$ and $b_{\ast}\approx b_{\mathrm{max}}$ at large $b$ values.
A typical value of $b_{\mathrm{max}}$ is chosen around $1\ \mathrm{GeV}^{-1}$, such that $b_{\ast}$ is always at the perturbative region.
There are several different $b_\ast$ prescriptions in the literature~\cite{Collins:2016hqq,Bacchetta:2017gcc}. In this work we adopt the original prescription introduced in Ref.~\cite{Collins:2011zzd}, with $b_{\ast}=b/\sqrt{1+b^2/b^2_{\mathrm{max}}}$.

In the small $b$ region $1/Q \ll b \ll 1/ \Lambda$, the defined TMDs at fixed energy $\mu$ can be expressed as a convolution of  perturbatively calculable hard coefficients and the corresponding collinear PDFs ~\cite{Collins:1981uk,Bacchetta:2013pqa}
\begin{equation}
f_{1\,q/H}(x,b;\mu)=\sum_i C_{q\leftarrow i}\otimes f_1^{i/H}(x,\mu),
\end{equation}
where $\otimes$ stands for the convolution in the momentum fraction $x$
\begin{equation}
 C_{q\leftarrow i}\otimes f_1^{i/H}(x,\mu)\equiv \int_{x}^1\frac{d\xi}{\xi} C_{q\leftarrow i}(x/\xi,b;\mu)f_1^{i/H}(\xi,\mu)
 \label{eq:otimes}
\end{equation}
and $f_1^{i/H}(\xi,\mu)$ is the corresponding collinear PDF of the $i$ flavor in hadron $H$ at the energy scale $\mu$, which can be a dynamic scale related to $b_*$ by $\mu_b=c_0/b_*$, with $c_0=2e^{-\gamma_E}$ and $\gamma_E\approx0.577$, the Euler Constant~\cite{Collins:1981uk}. In addition, the sum $\Sigma i$ runs over all parton flavors.
Independently on the type of initial hadrons, the perturbative hard coefficients have been calculated for the parton-target case~\cite{Collins:1981uw,Aybat:2011zv} and the results have been presented in Ref.~\cite{Bacchetta:2013pqa} (see also Appendix A of Ref.~\cite{Aybat:2011zv}) as
\begin{align}
\label{eq:cfactor}
&C_{q\leftarrow q^{\prime}}(x,\mu_b)=\delta_{qq^{\prime}}\left[\delta(1-x)+\frac{\alpha_s}{\pi}\left(\frac{C_F}{2}(1-x)\right)\right],\\
&C_{q\leftarrow g}(x,\mu_b)=\frac{\alpha_s}{\pi}T_Rx(1-x),
\end{align}
where $C_F=(N_C^2-1)/(2N_C)$, $T_R=1/2$, and $\alpha_s$ is the strong coupling constant at the energy scale $\mu_b$.
The $b_\ast$ prescription prevents $\alpha_s(\mu_b)$ from hitting the so-called Landau pole at large $b$ regime.
We note that the $C$-coefficients do not depend on the TMD scheme and on the initial hadrons. Namely, the $C$-coefficients in Eq.~(\ref{eq:cfactor}) are universal for both different schemes and initial hadrons.

The Sudakov-like form factor in Eq.~(\ref{eq:f}) can be separated into a perturbatively calculable part and a nonperturbative part
\begin{equation}
\label{eq:S}
S=S_{\mathrm{pert}}+S_{\mathrm{NP}}.
\end{equation}
The perturbative part of $S$ has the form
\begin{equation}
\label{eq:Spert}
S_{\mathrm{pert}}(Q,b)=\int^{Q^2}_{\mu_b^2}\frac{d\bar{\mu}^2}{\bar{\mu}^2}\left[A(\alpha_s(\bar{\mu}))
\mathrm{ln}\frac{Q^2}{\bar{\mu}^2}+B(\alpha_s(\bar{\mu}))\right].
\end{equation}
The coefficients $A$ and $B$ in Eq.(\ref{eq:Spert}) can be expanded as a $\alpha_s/{\pi}$ series:
\begin{align}
A=\sum_{n=1}^{\infty}A^{(n)}(\frac{\alpha_s}{\pi})^n,\\
B=\sum_{n=1}^{\infty}B^{(n)}(\frac{\alpha_s}{\pi})^n.
\end{align}
In this work, we take $A^{(n)}$ to $A^{(2)}$ and $B^{(n)}$ to $B^{(1)}$ up to the accuracy of next-to-leading-logarithmic (NLL) order~\cite{Collins:1984kg,Landry:2002ix,Qiu:2000ga,Kang:2011mr,Aybat:2011zv,Echevarria:2012pw}:
\begin{align}
A^{(1)}&=C_F\\
A^{(2)}&=\frac{C_F}{2}\left[C_A\left(\frac{67}{18}-\frac{\pi^2}{6}\right)-\frac{10}{9}T_Rn_f\right]\\
B^{(1)}&=-\frac{3}{2}C_F.
\end{align}

For the nonperturbative form factor $S_{\mathrm{NP}}$ in $pp$ Drell-Yan process, a general parameterization associated with the TMD PDF has been proposed in Ref.~\cite{Su:2014wpa} and it has the form
\begin{align}
S_{\mathrm{NP}}=g_1b^2+g_2\mathrm{ln}\frac{b}{b_{\ast}}\mathrm{ln}\frac{Q}{Q_0}+g_3b^2\left((x_0/x_1)^{\lambda}+(x_0/x_2)^\lambda\right).
\label{eq:SNP_DY_NN}
\end{align}
In Ref.~\cite{Su:2014wpa} the parameters $g_1,\ g_2,\ g_3$ are fitted from the nucleon-nucleon Drell-Yan process data
at the initial scale $Q^2_0=2.4\ \mathrm{GeV}^2$ with $b_{\mathrm{max}}=1.5\ \mathrm{GeV}^{-1}$, $x_0=0.01$ and $\lambda=0.2$.  With the extracted parameters, the fitted results are
$g_1=0.212,\ g_2=0.84, \ g_3 = 0$.

Since the nonperturbative form factor $S_{\mathrm{NP}}$ for quarks and antiquarks satisfies the following relation~\cite{Prokudin:2015ysa}
\begin{align}
S^q_{\mathrm{NP}}(Q,b)+S^{\bar{q}}_{\mathrm{NP}}(Q,b)=S_{\mathrm{NP}}(Q,b),
\end{align}
and the quark and antiquark contributions to $S_{\mathrm{NP}}$ are the same:
\begin{align}
S^q_{\mathrm{pert}}(Q,b_\ast)=S^{\bar{q}}_{\mathrm{pert}}(Q,b_\ast)=S_{\mathrm{pert}}(Q,b_\ast)/2,
\end{align}
the $S_{\mathrm{NP}}$ associated with the TMD distribution function of one of the initial protons can be expressed as
\begin{align}
\label{eq:SNPproton}
&S^{f_1^{q/p}}_{\mathrm{NP}}(Q,b)=\frac{g_1}{2}b^2+\frac{g_2}{2}\ln\frac{b}{b_{\ast}}\ln\frac{Q}{Q_0}.
\end{align}

With all the ingredients presented above, we can rewrite the  $b-$space unpolarized distribution function $f_1$ (e.g., for an up quark inside the proton), which depends on $x,b,Q$, as
\begin{align}
\label{eq:f1}
\tilde{f}_{1u/p}^\mathrm{sub}(x,b;Q)=e^{-\frac{1}{2}S_{\mathrm{pert}}(Q,b_\ast)-S^{f_1^{q/p}}_{\mathrm{NP}}(Q,b)}\mathcal{F}(\alpha_s(Q))
\sum_iC_{u\leftarrow i}\otimes f_{1}^{i/p}(x,\mu_b)
\end{align}
with $S_{\mathrm{pert}}(Q,b_\ast)$ and $S^{f_1^{q/p}}_{\mathrm{NP}}(Q,b)$ given in Eqs.(\ref{eq:Spert}) and (\ref{eq:SNPproton}).
Using Eq.(\ref{eq:cfactor}), we have
\begin{align}
&\sum_i C_{u\leftarrow i}\otimes f_{1}^{i/p}(x,\mu_b)\nonumber\\
&=C_{u\leftarrow q}\otimes f_1^{q/p}(x,\mu_b)+C_{u\leftarrow g}\otimes g(x,\mu_b)\nonumber\\
&=f_1^{u/p}(x,\mu_b)+\frac{\alpha_s}{\pi}\frac{C_F}{2}\int_{x}^1 \frac{d\xi}{\xi}(1-\frac{x}{\xi}) f_1^{u/p}(\xi,\mu_b)+
\frac{\alpha_s}{\pi}T_R\int_{x}^1\frac{d\xi}{\xi} \frac{x}{\xi}(1-\frac{x}{\xi})g(\xi,\mu_b),
\end{align}
where $g(x,\mu_b)$ stands for the distribution function of the gluon and $\mu_b$ turns to $\mu_b(b_{\ast})=c_0/b_{\ast}$.

If we perform a Fourier Transformation on $\tilde{f}_{1q/p}^\mathrm{sub}(x,b;Q)$, we can obtain the distribution function in transverse momentum space
\begin{align}
f_{1q/p}(x,k_\perp;Q)=\int_0^\infty\frac{dbb}{2\pi}J_0(k_\perp b)\tilde{f}_{1q/p}^\mathrm{sub}(x,b;Q),
\end{align}
where $J_0$ is the Bessel function of the first kind.

The nonperturbative Sudakov form factor for the pion TMD distributions has never been obtained. Here we assume that it has the same structure as for the proton TMD distributions, with parameters $g^{\pi}_1$ and $g^{\pi}_2$
\begin{align}
S^{f_1^{q/\pi}}_{\mathrm{NP}}=g^{\pi}_1b^2+g^{\pi}_2\mathrm{ln}\frac{b}{b_{\ast}}\mathrm{ln}\frac{Q}{Q_0}.
\label{eq:snppi}
\end{align}
Similarly, we can express the evolved result for the pion TMD distributions as
\begin{align}
\label{eq:f1pion}
\tilde{f}^\mathrm{sub}_{1q/\pi}(x,b;Q) &=e^{-\frac{1}{2}S_{\mathrm{pert}}(Q,b_\ast)-S^{f_1^{q/\pi}}_{\mathrm{NP}}(Q,b)}
\mathcal{F}(\alpha_s(Q))\sum_iC_{q\leftarrow i}\otimes f_{1}^{i/\pi}(x,\mu_b),
\end{align}
and also in transverse momentum space as
\begin{align}
f_{1q/\pi}(x,k_\perp;Q)=\int_0^\infty\frac{dbb}{2\pi}J_0(k_\perp b)\tilde{f}_{1q/\pi}^\mathrm{sub}(x,b;Q).
\end{align}

\subsection{Expression for the differential cross section}

We note that the hard factors $\mathcal{F}(\alpha_s(Q))$, which appears in Eq.(\ref{eq:f1}) and Eq.(\ref{eq:f1pion}), and $H_{UU}(Q;\mu)$ in Eq.~(\ref{eq:WUU_1}), are scheme dependent.
In the two different schemes, the Collins-11 scheme~\cite{Collins:2011zzd} and the Ji-Ma-Yuan scheme~\cite{Ji:2004wu,Ji:2004xq}, the hard factors $\mathcal{F}(\alpha_s(Q))$ and $H_{UU}(Q;\mu)$ have different forms:
\begin{align}
&\tilde{\mathcal{F}}^{\mathrm{JCC}}(\alpha_s(Q))=1+\mathcal{O}(\alpha_s^2), \\ &\tilde{\mathcal{F}}^{\mathrm{JMY}}(\alpha_s(Q),\rho)=1+\frac{\alpha_s}{2\pi}C_F
\left[\mathrm{ln}\rho-\frac{1}{2}\mathrm{ln}^2\rho-\frac{\pi^2}{2}-2\right], \\
&H^{\mathrm{JCC}}(Q;\mu)=1+\frac{\alpha_s(\mu)}{2\pi}C_F
\left(3\ln\frac{Q^2}{\mu^2}-\ln^2\frac{Q^2}{\mu^2}+\pi^2-8\right), \\
&H^{\mathrm{JMY}}(Q;\mu,\rho)=1+\frac{\alpha_s(\mu)}{2\pi}C_F
\left((1+\ln\rho^2)\ln\frac{Q^2}{\mu^2}-\ln\rho^2+\ln^2\rho+2\pi^2-4\right),
\end{align}
where $\rho$ is another parameter to regulate the light-cone singularity of TMD distributions.
However, if one absorbs both of the hard factors $H$ and $\mathcal{F}$ into the definition of the $C$-coefficients, the $C$-coefficients turn out to be $\rho$-independent (scheme-independent) and can be expresses as~\cite{Catani:2000vq}
\begin{align}
&C^\prime_{q\leftarrow q^{\prime}}(x,b;\mu_b)=\delta_{qq^{\prime}}\left[\delta(1-x)+\frac{\alpha_s}{\pi}
\left(\frac{C_F}{2}(1-x)+\frac{C_F}{4}(\pi^2-8)\delta(1-x)\right)\right],\label{eq:cfactor_ab1}\\
&C^\prime_{q\leftarrow g}(x,b;\mu_b)=\frac{\alpha_s}{\pi}T_Rx(1-x).
\label{eq:cfactor_ab2}
\end{align}
Substituting the above new $C$-coefficients into Eq.~(\ref{eq:f1}) and Eq.~(\ref{eq:f1pion}), we can obtain the evolved TMD distributions used in the calculation of the differential cross section for the unpolarized Drell-Yan process.

Finally, the structure function in $b$-space can be written as
\begin{align}
\label{eq:wuu}
\widetilde{W}_{UU}(Q;b)&=e^{-S_\mathrm{pert}(Q^2,b)-S^{f^{q/\pi}_{1}}_\mathrm{NP}(Q^2,b)-S^{f^{q/p}_{1}}_\mathrm{NP}(Q^2,b)}\nonumber\\
&\times \sum_{q,\bar{q}}e_q^2\,C^\prime_{q\leftarrow i}\otimes f_{i/\pi^-}(x_1,\mu_b)
\,C^\prime_{\bar{q}\leftarrow j}\otimes f_{j/p}(x_2,\mu_b).
\end{align}
After performing the Fourier Transformation, we can get the differential cross section, written in Eq.~(\ref{eq:dsigma}), as
\begin{align}
\label{eq:d4sigma}
&\frac{d^4\sigma}{dQ^2dyd^2\bm{q}_{\perp}}=\sigma_0\int_0^\infty \frac{db b}{2\pi}J_0(q_\perp b)\times \widetilde{W}_{UU}(Q;b).
\end{align}

\section{Extracting the nonperturbative Sudakov form factor for the pion TMD distribution}

\label{Sec.Numerical}

Using the framework set up above, in this section we fit the theoretical cross-section of $\pi^- N$ Drell-Yan at the kinematics of E615 measurement, to the corresponding experimental data~\cite{Conway:1989fs}, in order to extract the parameters of the Sudakov form factor for the pion TMD distributions.

\subsection{The differential cross section at E615 experiment}

The E615 experiments at FermiLab produced a $\pi^-$ beam with $P_\pi= 252$ GeV, which collided on a tungsten target (74 protons and 110 neutrons). The differential cross-section was given as 2-fold data depending on $x_F$ and $q_T$~\cite{Stirling:1993gc}, while the formula in Eq.~(\ref{eq:d4sigma}) denotes the 4-fold differential cross section as a function of $Q^2,\ y,\ \bm{q}_T$.
Thus, we apply the kinematics at E615
\begin{align}
0.2<x_\pi<1,\quad0.04<x_N<1,\quad0<x_F<1,\quad4.05\ \mathrm{GeV}<Q<8.55\ \mathrm{GeV},
\end{align}
to perform the integration over $Q^2$ from $4.05^2\ \mathrm{GeV}^2$ to $8.55^2\ \mathrm{GeV}^2$, and we transform the differential with respect to $x_F$ to arrive at
\begin{align}
\frac{d^2\sigma}{dx_Fdq_{\perp}}&=\sigma_0\frac{1}{\sqrt{x_F^2+4\frac{Q^2}{s}}}2\pi q_\perp\int_{4.05^2}^{8.55^2}dQ^2
\int_0^\infty \frac{db b}{2\pi}J_0(q_\perp b)\times \widetilde{W}_{UU}(Q;b).
\label{eq:dsigmadxdqt}
\end{align}

We adopt the NLO set of the CT10 parametrization~\cite{Lai:2010vv}~(central PDF sets) for the unpolarized distribution function $f_1(x)$ of the proton. For the unpolarized distribution function of the pion meson, we use the NLO SMRS parametrization~\cite{Sutton:1991ay}.
In our calculation we ignore any nuclear effects from the tungsten target.
The values of the strong coupling $\alpha_s$ are consistently obtained at 2-loop order as
\begin{align}
\alpha_s(Q^2)&=\frac{12\pi}{(33-2n_f)\mathrm{ln}(Q^2/\Lambda^2_{QCD})}
\left\{{1-\frac{6(153-19n_f)}{(33-2n_f)^2}
\frac{\mathrm{ln}\mathrm{ln}(Q^2/\Lambda^2_{QCD})}{\mathrm{ln}(Q^2/\Lambda^2_{QCD})}}\right\}, \label{eq:alphas}
\end{align}
with fixed $n_f=5$ and $\Lambda_{\mathrm{QCD}}=0.225\ \mathrm{GeV}$. We note that the running coupling in Eq.~(\ref{eq:alphas}) satisfies $\alpha_s(M_Z^2)=0.118$.

The experimental data that we use in the fit are the $q_\perp$-dependent differential cross sections of the unpolarized $\pi^- N$ Drell-Yan from the E615 experiment, as listed in Table.~36 of Ref.~\cite{Stirling:1993gc}.
The data span the region $0 <x_F<1$ with ten $x_F$ bins in a interval of 0.1. However, we find that a good fit can be obtained if we exclude the data in the bins $0.8<x_F<0.9$ and $0.9<x_F<1$.
Therefore, in the fit we disregard the data in the region $x_F>0.8$.
To estimate the result in each $x_F$ bin, we calculate the $q_\perp$-dependent cross section at the mean value of the $x_F$ bin.

\subsection{The fitting procedure}

The data fitting is performed by the package {\sc{MINUIT}}~\cite{James:1975dr,James:1994vla}, through a least-squares fit in which the chisquare is defined as
\begin{align}
\chi^2(\bm{\alpha})=\sum_{i=1}^{M}\sum_{j=1}^{N_i} \frac{(\mathrm{theo}(q_{\perp ij},\bm{\alpha})-\mathrm{data}_{ij})^2}{\mathrm{err}_{ij}^2}.
\label{eq:chisquare}
\end{align}
In the definition above, $i$ stands for the $i-$th $x_F$ bin out of the total of $M$ bins, each one with $N_i$ data points, $\bm{\alpha}$ is the vector of the free parameters: $g_1^\pi$ and $g_2^\pi$, $q_{\perp ij}$ denotes the $j-$th value of $q_\perp$ measured in the $i-$th $x_F$ bin, $\mathrm{theo}(q_{\perp ij},\bm{\alpha})$ is the theoretical estimate at $q_{\perp ij}$ and $\bm{\alpha}$, and $\mathrm{exp}_{ij}$ is the data measured at $q_{\perp ij}$ with error $\Delta \mathrm{err}_{ij}$.
The total number of data in our fit is $N=\sum_i^{8}N_i=96$.
We note that, apart from the statistical error in Table.~36 of Ref.~\cite{Stirling:1993gc}, there is also an overall systematic error of $16\%$.
Thus, in our fit the error is the total error which has the form $\Delta \mathrm{err}_{\mathrm{tot}}=\sqrt{\mathrm{err}_{\mathrm{stat}}^2+\mathrm{err}_{\mathrm{sys}}^2}$.
Since the TMD formalism is valid in the region $q_\perp\ll Q $, we do a simple data selection by removing the data in the region $q_\perp>3\ \mathrm{GeV}$.
\begin{figure}[H]
\centering
\includegraphics[width=0.8\columnwidth]{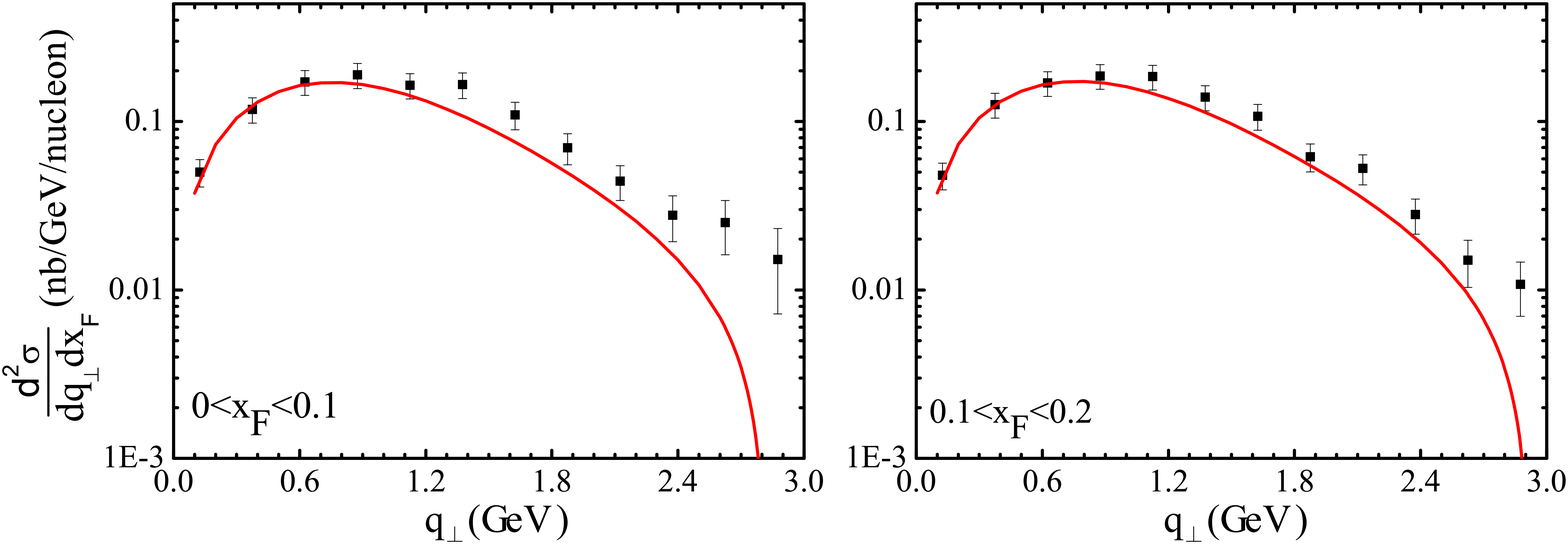}\\
\includegraphics[width=0.8\columnwidth]{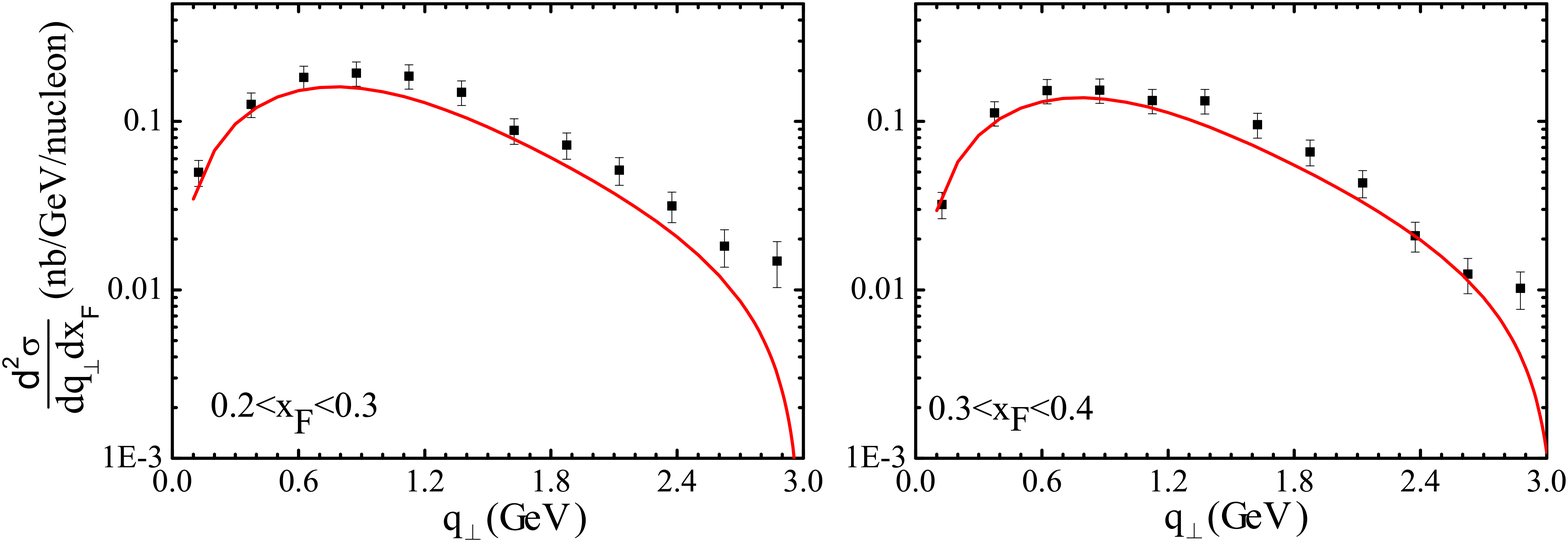}\\
\includegraphics[width=0.8\columnwidth]{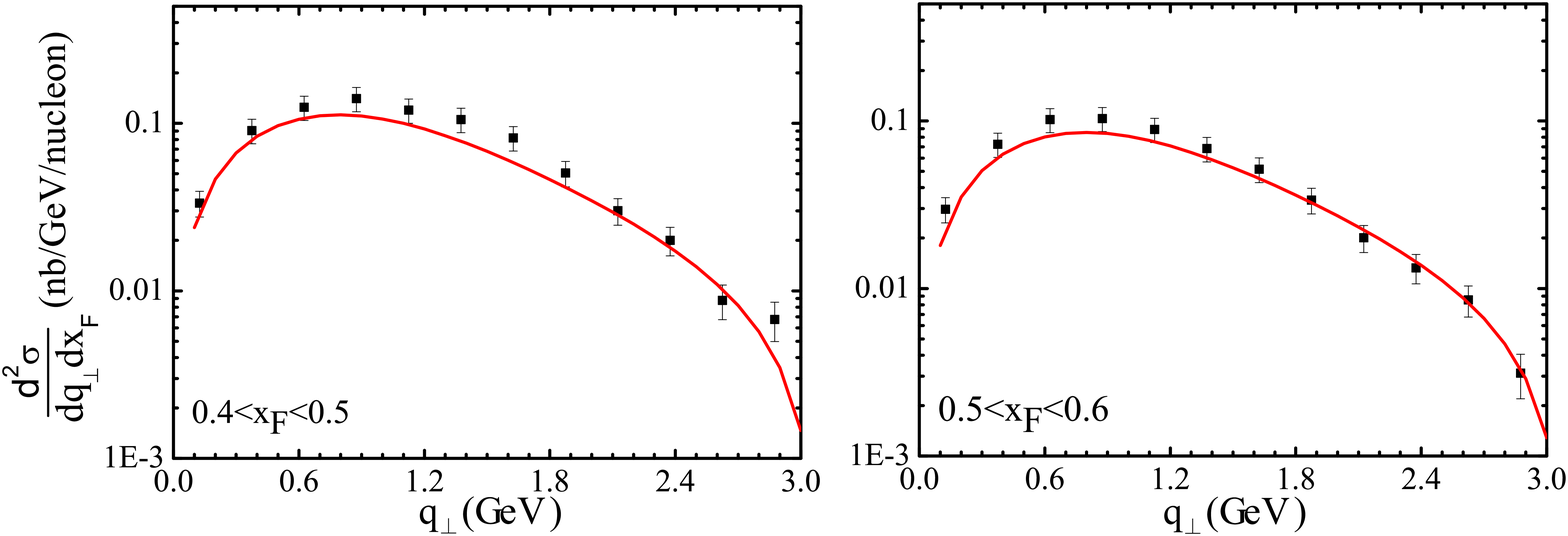}\\
\includegraphics[width=0.8\columnwidth]{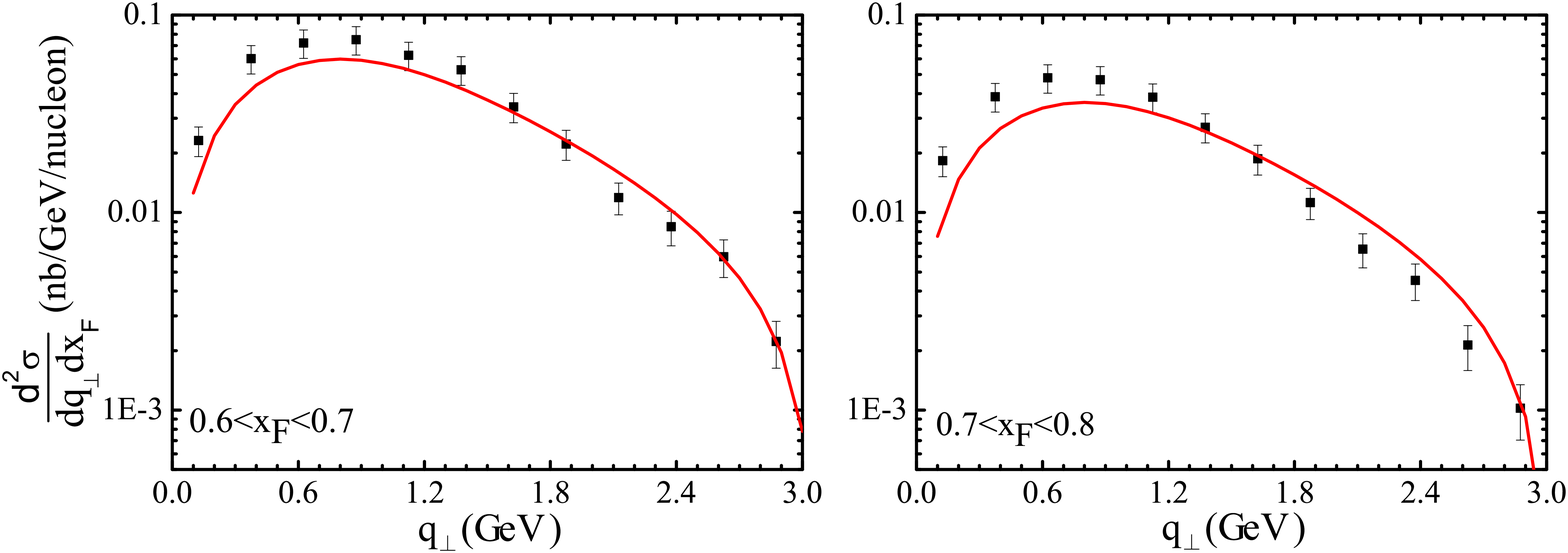}
\caption{The fitted cross section (solid line) of pion-nucleon Drell-Yan as functions of $q_\perp$, compared with the E615 data (full squre), for different $x_F$ bins in the range $0<x_F<0.8$. The error bars shown here include the statistical error and the $16\%$ systematic error.}
\label{fig:comparison1}
\end{figure}

We perform the fit by minimizing the chisquare in Eq.~(\ref{eq:chisquare}), and we obtain the following values for the two parameters:
\begin{align}
g^\pi_1=0.082 \pm 0.022 ,\quad g^\pi_2=0.394 \pm 0.103. \label{eq:gpi12}
\end{align}
with $\chi^2/\textrm{d.o.f} = 1.64$.
From the fitted result, we find that the values of the parameters $g_1$ and $g_2$ for the pion meson are smaller than those for the proton. Similar to the case of the proton, for the pion meson $g_2^\pi$ is several times larger than $g_1^\pi$.

In Fig.~\ref{fig:comparison1}, we plot the $q_\perp$-dependent differential cross section at different $x_F$ bins, using the fitted values for $g^\pi_1$ and $g^\pi_2$ in Eq.~(\ref{eq:gpi12}). The experimental data (full square) at E615 with error bars are also depicted for comparison.
As Fig.~\ref{fig:comparison1} demonstrates, a good fit is obtained at the region $x_F<0.8$.
We also check the theoretical result in the region $x_F>0.8$.
It turns out that, using the two parameters of the Sudakov form factor for the pion meson, the theoretical calculation underestimates the experimental data in that region, particularly when $x_F>0.9$.
The reason or this is that at large $x_F$ the higher twist effects dominates, and therefore the leading-twist TMD factorization breaks down.

\subsection{Results of TMDs for the pion at different scales}

\begin{figure}[H]
\includegraphics[width=0.98\columnwidth]{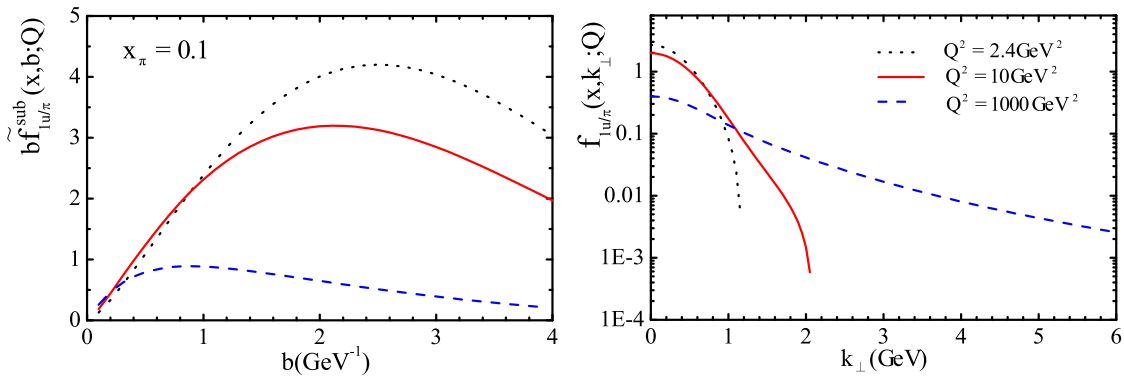}
\caption{Subtracted unpolarized TMD distribution of the pion meson for valence quarks in $b$-space (left panel) and $k_\perp$-space (right panel), at energies: $Q^2=2.4\ \mathrm{GeV}^2$~(dotted lines), $Q^2=10\ \mathrm{GeV}^2$~(solid lines) and $Q^2=1000\ \mathrm{GeV}^2$~(dashed lines).}
\label{fig:fsub}
\end{figure}

Using Eq.~(\ref{eq:f1pion}) and the extracted parameters $g_1^\pi$ and $g_2^\pi$, we quantitatively study the scale dependence of the pion TMD distributions.
In the left panel of Fig.~\ref{fig:fsub} we plot the subtracted valence distribution function in $b-$space $\tilde{f}_{1\,u/\pi^+}^{\textrm{sub}}(x,b;Q)$ (multiplied by $b$), for fixed $x_\pi=0.1$, at three different energy scales: $Q^2=2.4\ \mathrm{GeV}^2$ (dotted line), $10\ \mathrm{GeV}^2$ (solid line), $1000\ \mathrm{GeV}^2$ (dashed line).
In the right panel of Fig.~\ref{fig:fsub} we plot similar curves, but for the distribution $f_{1\,u/\pi^+}$ in transverse momentum space.
In this calculation, we have applied the $\rho$-independent $C-$coefficients presented in Eqs.~(\ref{eq:cfactor_ab1}) and (\ref{eq:cfactor_ab2}).
From the $b$-dependent plots, we see that at the highest energy scale $Q^2=1000\ \mathrm{GeV}^2$,  the peak of the curve is in the low $b$ region where $b<b_{\mathrm{max}}$, since in this case the perturbative part of the Sudakov form factor dominates.
However, at relatively lower energy scales, e.g., $Q^2=10\ \mathrm{GeV}^2$ and $Q^2=2.4\ \mathrm{GeV}^2$, the peak of the $b-$dependent distribution function moves towards the higher $b$ region, indicating that the nonperturbative part of the TMD evolution becomes more important at this lower energy scales.
For the distribution in transverse momentum $k_\perp$ space, at the high energy scale the distribution has a tail falling off slowly at large $k_\perp$,  while at lower energy scales the distribution function falls off rapidly with increasing $k_T$.
It is interesting to point out that the shapes of the pion TMD distribution at different scales are similar to those of the proton~\cite{Kang:2015msa}.

\section{Prediction of the $q_\perp$ spectrum of the lepton pair at COMPASS}
\label{Sec:COMPASS}

As a cross check, we calculate the $q_\perp$ spectrum of the lepton pair produced in the unpolarized pion-proton Drell-Yan process at the kinematics of COMPASS, using the TMD framework and applying the nonperturbative Sudakov form factor extracted in the previous section.
The COMPASS Collaboration recently reported the measurements on the transverse-spin-dependent azimuthal asymmetries with three different modulations in the $\pi^- N$ Drell-Yan process~\cite{Aghasyan:2017jop}, in which a $\pi^-$ beam with $P_\pi=\ 190\ \mathrm{GeV}$ collides on a $\mathrm{NH}_3$ target (10 protons and 7 neutrons).
These asymmetries provide a great opportunity to access the proton TMD distributions, such as the Sivers function, the transversity and prezelozity distributions, as well as the pion Boer-Mulders function.
Besides the measurement on the spin asymmetries, the COMPASS collaboration also presented the distribution of dimuon pairs as a function of the transverse momentum $q_\perp$ of the dimuon, up to $q_\perp =5.4$ GeV, which corresponds to the result in the case of an unpolarized target, since the azimuthal angles are integrated out.
The study of the unpolarized cross-section of $\pi N$ Drell-Yan is equally important, since it appears in the denominator of various asymmetries.
In particular, the study on the transverse momentum spectrum may also shed light on the unpolarized TMD distribution of the pion.

We use the following kinematics at COMPASS to calculate the $q_\perp$ distribution of the dimuon in $pi N$ Drell-Yan process
\begin{align}
&0.05<x_N<0.4,\quad 0.05<x_\pi<0.9, \quad
4.3\ \mathrm{GeV}<Q<8.5\ \mathrm{GeV}.
\label{eq:cuts}
\end{align}
\begin{figure}[H]
  \centering
  \includegraphics[width=0.6\columnwidth]{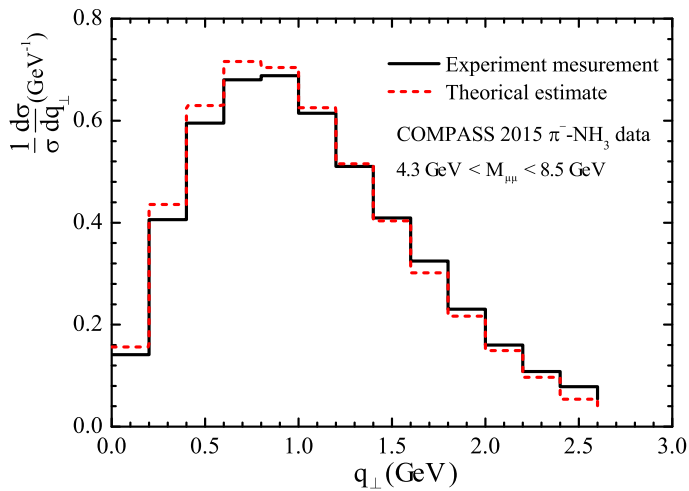}
  \caption{The transverse spectrum of lepton pair production in the unpolarized pion-nucleon Drell-Yan process, with an NH$_3$ target at COMPASS.
  The dashed line is our theoretical calculation using the extracted Sudakov form factor for the pion TMD PDF.
  The solid line shows the experimental measurement at COMPASS.}
  \label{fig:COMPASS}
\end{figure}
The differential cross section at COMPASS can be calculated via Eq.~(\ref{eq:dsigmadxdqt}) by replacing the integration limit of $Q$.
As shown in Ref.~\cite{Aghasyan:2017jop}, the events of the dimuon production were measured for different $q_\perp$ bins with an interval of $0.20\ \mathrm{GeV}$.
It can be also written as $N=\sigma\times\mathcal{L}$, where $\sigma$ is the total cross section and $\mathcal{L}$ is the integrated luminosity of incident hadrons.
To cancel the uncertainty of the integrated luminosity, we resort to the normalized differential cross section defined as~\cite{Khachatryan:2015oqa}
\begin{align}
\frac{1}{\sigma}\frac{d\sigma}{dq_\perp}=\frac{1}{\Sigma_i N_i}\frac{N_i}{\Delta q_\perp}.
\end{align}
Here, $N_i$ refers to the measured events at the $i-$th $q_\perp$ bin, and $\Delta q_\perp$ represents the width of $q_\perp$ bin.
For the cross section $\frac{d\sigma}{dq_\perp}$ as the function of $q_\perp$, we perform the integration over the entire $x_F$ region at COMPASS and estimate the value for different $q_\perp$ bins.
The total cross section $\sigma$ is calculated by integrating over $q_\perp$ and $x_F$.
To make the TMD factorization valid, we choose the upper limit of $q_\perp$ in our calculation to be $2.6\ \mathrm{GeV}$.

In Fig.~\ref{fig:COMPASS}, we plot the normalized transverse momentum spectrum of the dimuon pair produced in the pion-nucleon Drell-Yan process at COMPASS, at different $q_\perp$ bins, with an interval of 0.2 GeV.
The dashed line depicts our theoretical estimate which implements the extracted nonperturbative Sudakov form factor for the pion TMD distribution.
The solid line corresponds to the recent experimental measurement by the COMPASS collaboration~\cite{Aghasyan:2017jop}.
Comparing the two curves, we find that our theoretical estimate on the $q_\perp$ distribution of the dimuon agrees with the COMPASS data fairly well in the small $q_\perp$ region, in both the shape and the size.
In the region $q_\perp<2$ GeV, the difference between calculation and data is less then $10\%$.
This validates our extraction of the nonperturbative Sudakov form factor for the pion distribution $f_{1\pi}$, within the TMD factorization.
At relatively larger $q_\perp$ values, our calculation cannot describe the data, indicating that the perturbative correction from the $Y_{UU}$ term plays an important role in the region $q_\perp \sim Q$.
Further study is needed in order to provide a complete picture of the $q_\perp$ distribution of dilepton pairs from $\pi N$ Drell-Yan in the whole $q_\perp$ range.

\section{Conclusion}

\label{Sec.conclusion}

In this work, we applied the framework of the TMD factorization to study the transverse momentum distribution of the dilepton in the unpolarized $\pi^-N$ Drell-Yan process.
We took into account the TMD evolution of the distributions and considered the nonperturbative Sudakov form factor for the unpolarized TMD distribution function of the pion meson.
We parameterized the form factor in a form analogous to that of the proton TMD distribution, and we performed a fit
using the experimental data for the differential cross section as a function of $q_\perp$ at E615.
Adopting the extracted values of the parameters $g_1^\pi$ and $g_2^\pi$, we presented the evolved results of the unpolarized pion TMD distributions $f_{1\pi}$ at different energy scales.
We found a similarity between the TMD evolution of the pion distribution and that of the the proton distribution.
Finally, we calculated the transverse momentum spectrum of dimuon pairs produced in unpolarized pion-nucleon Drell-Yan processes at the COMPASS kinematics, and compared it with the COMPASS data to verify our extraction.
As a result, the extracted nonperturbative Sudakov form factor is compatible with the COMPASS measurement at small $q_\perp$ region with $q_\perp\ll Q$, indicating that our approach can be used as a first step to study the Drell-Yan process at COMPASS.
Our study may also provide a better understanding on the pion TMD distribution as well as its role in Drell-Yan process.
Furthermore, the framework applied in this work can also be extended to the study of the azimuthal asymmetries in the $\pi N$ Drell-Yan process, which we reserve for a future study.

\acknowledgments
This work is partially supported by the NSFC (China) grant 11575043, by the Fundamental Research Funds for the Central Universities of China, by Fondecyt (Chile) grants 1140390 and FB-0821. X. Wang is supported by the Scientific Research Foundation of Graduate School of Southeast University (Grants No.~YBJJ1667).

\bibliographystyle{JHEP}

\begin{thebibliography}{10}

\bibitem{Christenson:1970um}
J.~H. Christenson, G.~S. Hicks, L.~M. Lederman, P.~J. Limon, B.~G. Pope and
  E.~Zavattini, \emph{{Observation of massive muon pairs in hadron
  collisions}},
  \href{http://dx.doi.org/10.1103/PhysRevLett.25.1523}{\emph{Phys. Rev. Lett.}
  {\bfseries 25} (1970) 1523--1526}.

\bibitem{Drell:1970wh}
S.~D. Drell and T.-M. Yan, \emph{{Massive Lepton Pair Production in
  Hadron-Hadron Collisions at High-Energies}},
  \href{http://dx.doi.org/10.1103/PhysRevLett.25.316}{\emph{Phys. Rev. Lett.}
  {\bfseries 25} (1970) 316--320}.

\bibitem{Bordalo:1987cs}
{\scshape NA10} collaboration, P.~Bordalo et~al., \emph{{Nuclear Effects on the
  Nucleon Structure Functions in Hadronic High Mass Dimuon Production}},
  \href{http://dx.doi.org/10.1016/0370-2693(87)91253-6}{\emph{Phys. Lett.}
  {\bfseries B193} (1987) 368}.

\bibitem{Betev:1985pf}
{\scshape NA10} collaboration, B.~Betev et~al., \emph{{Differential
  Cross-section of High Mass Muon Pairs Produced by a 194-{GeV}/$c \pi^-$ Beam
  on a Tungsten Target}}, \href{http://dx.doi.org/10.1007/BF01550243}{\emph{Z.
  Phys.} {\bfseries C28} (1985) 9}.

\bibitem{Falciano:1986wk}
{\scshape NA10} collaboration, S.~Falciano et~al., \emph{{Angular Distributions
  of Muon Pairs Produced by 194-{GeV}/$c$ Negative Pions}},
  \href{http://dx.doi.org/10.1007/BF01551072}{\emph{Z. Phys.} {\bfseries C31}
  (1986) 513}.

\bibitem{Guanziroli:1987rp}
{\scshape NA10} collaboration, M.~Guanziroli et~al., \emph{{Angular
  Distributions of Muon Pairs Produced by Negative Pions on Deuterium and
  Tungsten}}, \href{http://dx.doi.org/10.1007/BF01549713}{\emph{Z. Phys.}
  {\bfseries C37} (1988) 545}.

\bibitem{Conway:1989fs}
J.~S. Conway et~al., \emph{{Experimental Study of Muon Pairs Produced by
  252-GeV Pions on Tungsten}},
  \href{http://dx.doi.org/10.1103/PhysRevD.39.92}{\emph{Phys. Rev.} {\bfseries
  D39} (1989) 92--122}.

\bibitem{Palestini:1985zc}
S.~Palestini et~al., \emph{{Pion Structure as Observed in the Reaction $\pi^- N
  \to \mu^+ \mu^-$ X at 80-{GeV}/$c$}},
  \href{http://dx.doi.org/10.1103/PhysRevLett.55.2649}{\emph{Phys. Rev. Lett.}
  {\bfseries 55} (1985) 2649}.

\bibitem{Anassontzis:1987hk}
E.~Anassontzis et~al., \emph{{High mass dimuon production in $\bar{p} n$ and
  $\pi^- n$ interactions at 125-GeV/c}},
  \href{http://dx.doi.org/10.1103/PhysRevD.38.1377}{\emph{Phys. Rev.}
  {\bfseries D38} (1988) 1377}.

\bibitem{Gautheron:2010wva}
{\scshape COMPASS} collaboration, F.~Gautheron et~al.,
\href{http://wwwcompass.cern.ch/compass/proposal/compass-II_proposal/compass-II_proposal.pdf}{\emph{COMPASS-II
  Proposal}}.

\bibitem{Aghasyan:2017jop}
M.~Aghasyan et~al., \emph{{First measurement of transverse-spin-dependent
  azimuthal asymmetries in the Drell-Yan process}},
  \href{https://arxiv.org/abs/1704.00488}{{\ttfamily 1704.00488}}.

\bibitem{Brodsky:2002cx}
S.~J. Brodsky, D.~S. Hwang and I.~Schmidt, \emph{{Final state interactions and
  single spin asymmetries in semiinclusive deep inelastic scattering}},
  \href{http://dx.doi.org/10.1016/S0370-2693(02)01320-5}{\emph{Phys. Lett.}
  {\bfseries B530} (2002) 99--107},
  [\href{https://arxiv.org/abs/hep-ph/0201296}{{\ttfamily hep-ph/0201296}}].

\bibitem{Brodsky:2002rv}
S.~J. Brodsky, D.~S. Hwang and I.~Schmidt, \emph{{Initial state interactions
  and single spin asymmetries in Drell-Yan processes}},
  \href{http://dx.doi.org/10.1016/S0550-3213(02)00617-X}{\emph{Nucl. Phys.}
  {\bfseries B642} (2002) 344--356},
  [\href{https://arxiv.org/abs/hep-ph/0206259}{{\ttfamily hep-ph/0206259}}].

\bibitem{Collins:2002kn}
J.~C. Collins, \emph{{Leading twist single transverse-spin asymmetries:
  Drell-Yan and deep inelastic scattering}},
  \href{http://dx.doi.org/10.1016/S0370-2693(02)01819-1}{\emph{Phys. Lett.}
  {\bfseries B536} (2002) 43--48},
  [\href{https://arxiv.org/abs/hep-ph/0204004}{{\ttfamily hep-ph/0204004}}].

\bibitem{Anselmino:2009st}
M.~Anselmino, M.~Boglione, U.~D'Alesio, S.~Melis, F.~Murgia and A.~Prokudin,
  \emph{{Sivers effect in Drell-Yan processes}},
  \href{http://dx.doi.org/10.1103/PhysRevD.79.054010}{\emph{Phys. Rev.}
  {\bfseries D79} (2009) 054010},
  [\href{https://arxiv.org/abs/0901.3078}{{\ttfamily 0901.3078}}].

\bibitem{Kang:2009bp}
Z.-B. Kang and J.-W. Qiu, \emph{{Testing the Time-Reversal Modified
  Universality of the Sivers Function}},
  \href{http://dx.doi.org/10.1103/PhysRevLett.103.172001}{\emph{Phys. Rev.
  Lett.} {\bfseries 103} (2009) 172001},
  [\href{https://arxiv.org/abs/0903.3629}{{\ttfamily 0903.3629}}].

\bibitem{Peng:2014hta}
J.-C. Peng and J.-W. Qiu, \emph{{Novel phenomenology of parton distributions
  from the Drell¨CYan process}},
  \href{http://dx.doi.org/10.1016/j.ppnp.2014.01.005}{\emph{Prog. Part. Nucl.
  Phys.} {\bfseries 76} (2014) 43--75},
  [\href{https://arxiv.org/abs/1401.0934}{{\ttfamily 1401.0934}}].

\bibitem{Echevarria:2014xaa}
M.~G. Echevarria, A.~Idilbi, Z.-B. Kang and I.~Vitev, \emph{{QCD Evolution of
  the Sivers Asymmetry}},
  \href{http://dx.doi.org/10.1103/PhysRevD.89.074013}{\emph{Phys. Rev.}
  {\bfseries D89} (2014) 074013},
  [\href{https://arxiv.org/abs/1401.5078}{{\ttfamily 1401.5078}}].

\bibitem{Huang:2015vpy}
J.~Huang, Z.-B. Kang, I.~Vitev and H.~Xing, \emph{{Spin asymmetries for vector
  boson production in polarized p+p collisions}},
  \href{http://dx.doi.org/10.1103/PhysRevD.93.014036}{\emph{Phys. Rev.}
  {\bfseries D93} (2016) 014036},
  [\href{https://arxiv.org/abs/1511.06764}{{\ttfamily 1511.06764}}].

\bibitem{Anselmino:2016uie}
M.~Anselmino, M.~Boglione, U.~D'Alesio, F.~Murgia and A.~Prokudin, \emph{{Study
  of the sign change of the Sivers function from STAR Collaboration W/Z
  production data}},
  \href{http://dx.doi.org/10.1007/JHEP04(2017)046}{\emph{JHEP} {\bfseries 04}
  (2017) 046}, [\href{https://arxiv.org/abs/1612.06413}{{\ttfamily
  1612.06413}}].

\bibitem{Fermilab1} L.D.~Isenhower et~al.,
\href{http://www.fnal.gov/directorate/program_planning/June2012Public/P-1027_Pol-Drell-Yan-proposal.pdf}
{\emph{Fermilab DY proposal- polarized beam}}.

\bibitem{Fermilab2} D.~Geesaman et~al.,
\href{http://www.fnal.gov/directorate/program_planning/June2013PACPublic/P-1039_LOI_polarized_DY.pdf}
{\emph{Fermilab DY proposal- polarized target}}.

\bibitem{ANDY} E.C.~Aschenauer et~al., \href{https://www.bnl.gov/npp/docs/pac0611/DY_pro_110516_final.2.pdf}
{\emph{ANDY proposal}}.


\bibitem{Adamczyk:2015gyk}
{\scshape STAR} collaboration, L.~Adamczyk et~al., \emph{{Measurement of the
  transverse single-spin asymmetry in $p^\uparrow+p \to W^{\pm}/Z^0$ at RHIC}},
  \href{http://dx.doi.org/10.1103/PhysRevLett.116.132301}{\emph{Phys. Rev.
  Lett.} {\bfseries 116} (2016) 132301},
  [\href{https://arxiv.org/abs/1511.06003}{{\ttfamily 1511.06003}}].

\bibitem{Adolph:2016dvl}
{\scshape COMPASS} collaboration, C.~Adolph et~al., \emph{{Sivers asymmetry
  extracted in SIDIS at the hard scales of the Drell¨CYan process at COMPASS}},
  \href{http://dx.doi.org/10.1016/j.physletb.2017.04.042}{\emph{Phys. Lett.}
  {\bfseries B770} (2017) 138--145},
  [\href{https://arxiv.org/abs/1609.07374}{{\ttfamily 1609.07374}}].

\bibitem{Collins:1981uk}
J.~C. Collins and D.~E. Soper, \emph{{Back-To-Back Jets in QCD}},
  \href{http://dx.doi.org/10.1016/0550-3213(81)90339-4}{\emph{Nucl. Phys.}
  {\bfseries B193} (1981) 381}.

\bibitem{Collins:2011zzd}
J.~Collins, \emph{{Foundations of perturbative QCD}}.
\newblock Cambridge University Press, 2013.

\bibitem{Ji:2004wu}
X.-d. Ji, J.-p. Ma and F.~Yuan, \emph{{QCD factorization for semi-inclusive
  deep-inelastic scattering at low transverse momentum}},
  \href{http://dx.doi.org/10.1103/PhysRevD.71.034005}{\emph{Phys. Rev.}
  {\bfseries D71} (2005) 034005},
  [\href{https://arxiv.org/abs/hep-ph/0404183}{{\ttfamily hep-ph/0404183}}].

\bibitem{Aybat:2011zv}
S.~M. Aybat and T.~C. Rogers, \emph{{TMD Parton Distribution and Fragmentation
  Functions with QCD Evolution}},
  \href{http://dx.doi.org/10.1103/PhysRevD.83.114042}{\emph{Phys. Rev.}
  {\bfseries D83} (2011) 114042},
  [\href{https://arxiv.org/abs/1101.5057}{{\ttfamily 1101.5057}}].

\bibitem{Collins:2012uy}
J.~C. Collins and T.~C. Rogers, \emph{{Equality of Two Definitions for
  Transverse Momentum Dependent Parton Distribution Functions}},
  \href{http://dx.doi.org/10.1103/PhysRevD.87.034018}{\emph{Phys. Rev.}
  {\bfseries D87} (2013) 034018},
  [\href{https://arxiv.org/abs/1210.2100}{{\ttfamily 1210.2100}}].

\bibitem{Ji:2002aa}
X.-d. Ji and F.~Yuan, \emph{{Parton distributions in light cone gauge: Where
  are the final state interactions?}},
  \href{http://dx.doi.org/10.1016/S0370-2693(02)02384-5}{\emph{Phys. Lett.}
  {\bfseries B543} (2002) 66--72},
  [\href{https://arxiv.org/abs/hep-ph/0206057}{{\ttfamily hep-ph/0206057}}].

\bibitem{Echevarria:2012pw}
M.~G. Echevarria, A.~Idilbi, A.~Sch\"{a}fer and I.~Scimemi,
  \emph{{Model-Independent Evolution of Transverse Momentum Dependent
  Distribution Functions (TMDs) at NNLL}},
  \href{http://dx.doi.org/10.1140/epjc/s10052-013-2636-y}{\emph{Eur. Phys. J.}
  {\bfseries C73} (2013) 2636},
  [\href{https://arxiv.org/abs/1208.1281}{{\ttfamily 1208.1281}}].
\bibitem{Pitonyak:2013dsu}
D.~Pitonyak, M.~Schlegel and A.~Metz, \emph{{Polarized hadron pair production
  from electron-positron annihilation}},
  \href{http://dx.doi.org/10.1103/PhysRevD.89.054032}{\emph{Phys. Rev.}
  {\bfseries D89} (2014) 054032},
  [\href{https://arxiv.org/abs/1310.6240}{{\ttfamily 1310.6240}}].

\bibitem{Boer:2008fr}
D.~Boer, \emph{{Angular dependences in inclusive two-hadron production at
  BELLE}}, \href{http://dx.doi.org/10.1016/j.nuclphysb.2008.06.011}{\emph{Nucl.
  Phys.} {\bfseries B806} (2009) 23--67},
  [\href{https://arxiv.org/abs/0804.2408}{{\ttfamily 0804.2408}}].
\bibitem{Boer:2006eq}
D.~Boer and W.~Vogelsang, \emph{{Drell-Yan lepton angular distribution at small
  transverse momentum}},
  \href{http://dx.doi.org/10.1103/PhysRevD.74.014004}{\emph{Phys. Rev.}
  {\bfseries D74} (2006) 014004},
  [\href{https://arxiv.org/abs/hep-ph/0604177}{{\ttfamily hep-ph/0604177}}].

\bibitem{Arnold:2008kf}
S.~Arnold, A.~Metz and M.~Schlegel, \emph{{Dilepton production from polarized
  hadron hadron collisions}},
  \href{http://dx.doi.org/10.1103/PhysRevD.79.034005}{\emph{Phys. Rev.}
  {\bfseries D79} (2009) 034005},
  [\href{https://arxiv.org/abs/0809.2262}{{\ttfamily 0809.2262}}].

\bibitem{Collins:1984kg}
J.~C. Collins, D.~E. Soper and G.~F. Sterman, \emph{{Transverse Momentum
  Distribution in Drell-Yan Pair and W and Z Boson Production}},
  \href{http://dx.doi.org/10.1016/0550-3213(85)90479-1}{\emph{Nucl. Phys.}
  {\bfseries B250} (1985) 199--224}.

\bibitem{Lambertsen:2016wgj}
M.~Lambertsen and W.~Vogelsang, \emph{{Drell-Yan lepton angular distributions
  in perturbative QCD}},
  \href{http://dx.doi.org/10.1103/PhysRevD.93.114013}{\emph{Phys. Rev.}
  {\bfseries D93} (2016) 114013},
  [\href{https://arxiv.org/abs/1605.02625}{{\ttfamily 1605.02625}}].
\bibitem{Ji:2004xq}
X.-d. Ji, J.-P. Ma and F.~Yuan, \emph{{QCD factorization for spin-dependent
  cross sections in DIS and Drell-Yan processes at low transverse momentum}},
  \href{http://dx.doi.org/10.1016/j.physletb.2004.07.026}{\emph{Phys. Lett.}
  {\bfseries B597} (2004) 299--308},
  [\href{https://arxiv.org/abs/hep-ph/0405085}{{\ttfamily hep-ph/0405085}}].

\bibitem{Collins:1999dz}
J.~C. Collins and F.~Hautmann, \emph{{Infrared divergences and nonlightlike
  eikonal lines in Sudakov processes}},
  \href{http://dx.doi.org/10.1016/S0370-2693(99)01384-2}{\emph{Phys. Lett.}
  {\bfseries B472} (2000) 129--134},
  [\href{https://arxiv.org/abs/hep-ph/9908467}{{\ttfamily hep-ph/9908467}}].

\bibitem{Prokudin:2015ysa}
A.~Prokudin, P.~Sun and F.~Yuan, \emph{{Scheme dependence and transverse
  momentum distribution interpretation of Collins-Soper-Sterman resummation}},
  \href{http://dx.doi.org/10.1016/j.physletb.2015.09.064}{\emph{Phys. Lett.}
  {\bfseries B750} (2015) 533--538},
  [\href{https://arxiv.org/abs/1505.05588}{{\ttfamily 1505.05588}}].

\bibitem{Collins:2016hqq}
J.~Collins, L.~Gamberg, A.~Prokudin, T.~C. Rogers, N.~Sato and B.~Wang,
  \emph{{Relating Transverse Momentum Dependent and Collinear Factorization
  Theorems in a Generalized Formalism}},
  \href{http://dx.doi.org/10.1103/PhysRevD.94.034014}{\emph{Phys. Rev.}
  {\bfseries D94} (2016) 034014},
  [\href{https://arxiv.org/abs/1605.00671}{{\ttfamily 1605.00671}}].

\bibitem{Bacchetta:2017gcc}
A.~Bacchetta, F.~Delcarro, C.~Pisano, M.~Radici and A.~Signori,
  \emph{{Extraction of partonic transverse momentum distributions from
  semi-inclusive deep-inelastic scattering, Drell-Yan and Z-boson production}},
  \href{http://dx.doi.org/10.1007/JHEP06(2017)081}{\emph{JHEP} {\bfseries 06}
  (2017) 081}, [\href{https://arxiv.org/abs/1703.10157}{{\ttfamily
  1703.10157}}].

\bibitem{Bacchetta:2013pqa}
A.~Bacchetta and A.~Prokudin, \emph{{Evolution of the helicity and transversity
  Transverse-Momentum-Dependent parton distributions}},
  \href{http://dx.doi.org/10.1016/j.nuclphysb.2013.07.013}{\emph{Nucl. Phys.}
  {\bfseries B875} (2013) 536--551},
  [\href{https://arxiv.org/abs/1303.2129}{{\ttfamily 1303.2129}}].

\bibitem{Collins:1981uw}
J.~C. Collins and D.~E. Soper, \emph{{Parton Distribution and Decay
  Functions}},
  \href{http://dx.doi.org/10.1016/0550-3213(82)90021-9}{\emph{Nucl. Phys.}
  {\bfseries B194} (1982) 445--492}.

\bibitem{Landry:2002ix}
F.~Landry, R.~Brock, P.~M. Nadolsky and C.~P. Yuan, \emph{{Tevatron Run-1 $Z$
  boson data and Collins-Soper-Sterman resummation formalism}},
  \href{http://dx.doi.org/10.1103/PhysRevD.67.073016}{\emph{Phys. Rev.}
  {\bfseries D67} (2003) 073016},
  [\href{https://arxiv.org/abs/hep-ph/0212159}{{\ttfamily hep-ph/0212159}}].

\bibitem{Qiu:2000ga}
J.-w. Qiu and X.-f. Zhang, \emph{{QCD prediction for heavy boson transverse
  momentum distributions}},
  \href{http://dx.doi.org/10.1103/PhysRevLett.86.2724}{\emph{Phys. Rev. Lett.}
  {\bfseries 86} (2001) 2724--2727},
  [\href{https://arxiv.org/abs/hep-ph/0012058}{{\ttfamily hep-ph/0012058}}].

\bibitem{Kang:2011mr}
Z.-B. Kang, B.-W. Xiao and F.~Yuan, \emph{{QCD Resummation for Single Spin
  Asymmetries}},
  \href{http://dx.doi.org/10.1103/PhysRevLett.107.152002}{\emph{Phys. Rev.
  Lett.} {\bfseries 107} (2011) 152002},
  [\href{https://arxiv.org/abs/1106.0266}{{\ttfamily 1106.0266}}].

\bibitem{Su:2014wpa}
P.~Sun, J.~Isaacson, C.~P. Yuan and F.~Yuan, \emph{{Universal Non-perturbative
  Functions for SIDIS and Drell-Yan Processes}},
  \href{https://arxiv.org/abs/1406.3073}{{\ttfamily 1406.3073}}.

\bibitem{Catani:2000vq}
S.~Catani, D.~de~Florian and M.~Grazzini, \emph{{Universality of nonleading
  logarithmic contributions in transverse momentum distributions}},
  \href{http://dx.doi.org/10.1016/S0550-3213(00)00617-9}{\emph{Nucl. Phys.}
  {\bfseries B596} (2001) 299--312},
  [\href{https://arxiv.org/abs/hep-ph/0008184}{{\ttfamily hep-ph/0008184}}].

\bibitem{Stirling:1993gc}
W.~J. Stirling and M.~R. Whalley, \emph{{A Compilation of Drell-Yan
  cross-sections}},
  \href{http://dx.doi.org/10.1088/0954-3899/19/D/001}{\emph{J. Phys.}
  {\bfseries G19} (1993) D1--D102}.

\bibitem{Lai:2010vv}
H.-L. Lai, M.~Guzzi, J.~Huston, Z.~Li, P.~M. Nadolsky, J.~Pumplin et~al.,
  \emph{{New parton distributions for collider physics}},
  \href{http://dx.doi.org/10.1103/PhysRevD.82.074024}{\emph{Phys. Rev.}
  {\bfseries D82} (2010) 074024},
  [\href{https://arxiv.org/abs/1007.2241}{{\ttfamily 1007.2241}}].

\bibitem{Sutton:1991ay}
P.~J. Sutton, A.~D. Martin, R.~G. Roberts and W.~J. Stirling, \emph{{Parton
  distributions for the pion extracted from Drell-Yan and prompt photon
  experiments}}, \href{http://dx.doi.org/10.1103/PhysRevD.45.2349}{\emph{Phys.
  Rev.} {\bfseries D45} (1992) 2349--2359}.

\bibitem{James:1975dr}
F.~James and M.~Roos, \emph{{Minuit: A System for Function Minimization and
  Analysis of the Parameter Errors and Correlations}},
  \href{http://dx.doi.org/10.1016/0010-4655(75)90039-9}{\emph{Comput. Phys.
  Commun.} {\bfseries 10} (1975) 343--367}.

\bibitem{James:1994vla}
F.~James, \emph{{MINUIT Function Minimization and Error Analysis: Reference
  Manual Version 94.1}}, .

\bibitem{Kang:2015msa}
Z.-B. Kang, A.~Prokudin, P.~Sun and F.~Yuan, \emph{{Extraction of Quark
  Transversity Distribution and Collins Fragmentation Functions with QCD
  Evolution}}, \href{http://dx.doi.org/10.1103/PhysRevD.93.014009}{\emph{Phys.
  Rev.} {\bfseries D93} (2016) 014009},
  [\href{https://arxiv.org/abs/1505.05589}{{\ttfamily 1505.05589}}].

\bibitem{Khachatryan:2015oqa}
{\scshape CMS} collaboration, V.~Khachatryan et~al., \emph{{Measurement of the
  differential cross section for top quark pair production in pp collisions at
  $\sqrt{s} = 8\,\text {TeV} $}},
  \href{http://dx.doi.org/10.1140/epjc/s10052-015-3709-x}{\emph{Eur. Phys. J.}
  {\bfseries C75} (2015) 542},
  [\href{https://arxiv.org/abs/1505.04480}{{\ttfamily 1505.04480}}].






\end{thebibliography}

\end{document}